\documentclass[aps,prl,twocolumn,groupedaddress]{revtex4-1}

\usepackage{graphicx}
\usepackage{dcolumn}
\usepackage[dvips]{color} 

\begin{document}


\title{Fermi Acceleration in Plasmoids interacting with Fast Shocks of Reconnection via Fractal Reconnection}


\author{Naoto Nishizuka$^1$}
\author{Kazunari Shibata$^2$}
\affiliation{%
$^1$Institute of Space and Astronautical Science, Japan Aerospace Exploration Agency, Sagamihara, Kanagawa, 252-5210, Japan\\
$^2$Kwasan and Hida observatories, Kyoto University, Kyoto, 607-8471, Japan
}%


\date{\today}

\begin{abstract}
We propose the particle acceleration model coupled with multiple plasmoid ejections in a solar flare. Unsteady reconnection produces plasmoids in a current 
sheet and ejects them out to the fast shocks, where particles in a plasmoid are reflected upstream the shock front by magnetic mirror effect. As the plasmoid 
passes through the shock front, the reflection distance becomes shorter and shorter driving Fermi acceleration, until it becomes proton Larmor radius. The 
fractal distribution of plasmoids may also have a role in naturally explaining the power-law spectrum in nonthermal emissions.
\end{abstract}


\maketitle

%
{\it Introduction} --
Recent observations of solar X-ray, gamma-ray and microwave bursts revealed that energy release in a solar flare is very dynamic and that high energy particles, i.e. 
GeV ions and MeV electrons, are generated in a very short time period ($<$1 s). This short time variability of bursts, sometimes fractal-like time variability, indicates 
highly fragmented acceleration regions \cite{asc01}, and these are expected to be above or around the loop-top hard X-ray (HXR) source \cite{mas94, asc96}. HXR 
spectral observations of solar flares have established that efficient electron acceleration (10$^{34}$-10$^{35}$ electrons/s) occurs during impulsive phase of solar 
flares \cite{lin71}. To explain the high energy particles, several models have been considered, such as DC field acceleration inside a current sheet \cite{hol85, ben94, 
lit96}, stochastic acceleration in the turbulent reconnection outflow \cite{bro85, ben87, ter89, mil96} and shock acceleration at the fast shock \cite{som97, tsu98}, 
though assumed turbulent flows are still not revealed. Furthermore, recent studies have shown that the role of multiple X-points in a current sheet is more important 
for particle acceleration and energy release \cite{dra06, dau11, hos12}.

\begin{figure}[hbp]
\includegraphics{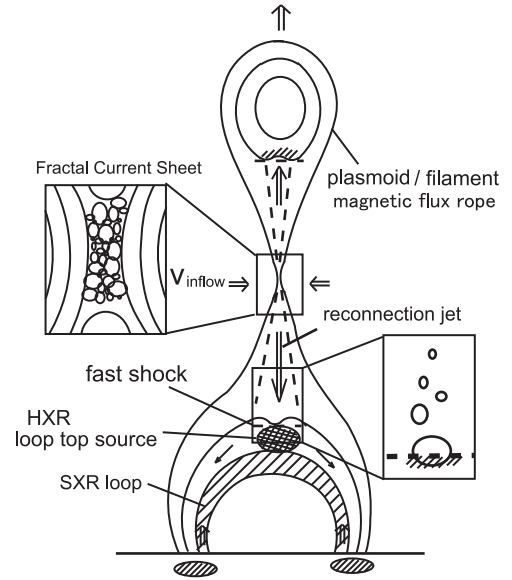}
\caption{\label{fig1} Particle acceleration driven by multiple plasmoid ejections colliding with the fast shock. Multiple plasmoids of 
various scales are intermittently ejected upward and downward out of a turbulent current sheet and collide with the termination 
shocks of reconnection outflows, i.e. fast shocks above and below a reconnection X-point, where particles are effectively accelerated 
via shock acceleration process trapped in a plasmoid.}
\end{figure}

X-ray emitting plasma ejection, shortly plasmoid ejection, is one of the direct evidence of magnetic reconnection in a solar flare \cite{shi11}. In the pre-flare phase, 
a plasmoid (a magnetic flux rope) is observed to gradually rise up, until it is accelerated upward in the impulsive phase in association with HXR burst \cite{ohy97, tsu98}. 
Multiple plasmoid ejection events have been discovered in radio and soft X-ray (SXR) observations in association with HXR bursts \cite{kar04, nis10}. It is also reported 
that multiple downflows, which are counterparts of multiple ejections, associate with HXR bursts \cite{asa04}. These observations may indicate the relationship between 
particle acceleration and plasmoid ejections. Furthermore, some plasmoids were observed falling down to the loop-top in UV and HXR images \cite{kol07, sui03}.

Plasmoid ejection drives dynamic feature of magnetic reconnection. It is known that plasmoid ejection induces inflow into the current sheet and increases reconnection 
rate \cite{shi01, nisd09}. Plasmoid formation repeats self-similarly in a current sheet and makes fractal-like or turbulent structure in a current sheet via fractal reconnection 
\cite{shi01, hua10, lou09, sam09, bar11, kow11, lou11, she11}. Plasmoids in a current sheet are unstable for coalescence instability and repeat lots of collisions with each other, then 
merging to a single large plasmoid which is finally ejected out of the current sheet. During this process, strong DC-field is enhanced between the two colliding plasmoids and 
shrinkage of a large plasmoid also accelerates particles impulsively, until a largest plasmoid is ejected outside of the current sheet \cite{taj87, dra06, oka10, tan10, kar11}. 
The large numbers of plasmoids remain as the exhaust impacts the loop-top shock. This is directly observed as multiple plasmoid ejections and downflows correlated with 
HXR emission \cite{asa04, nis10} and indirectly as power-law distributions of HXR and UV footpoint bright points \cite{ben94, nis09}.

\begin{figure}[htbp]
\includegraphics{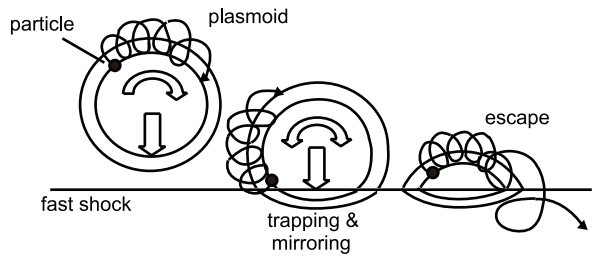}
\caption{\label{fig2} Scenario of shock acceleration at the fast shock trapped in a plasmoid; particles trapped in a plasmoid collide with the fast shock, 
when they repeat reflections upstream of the shock because of magnetic pressure gradient (mirror force). During the passage of a plasmoid through 
the shock front, trapping distance of particles becomes shorter and shorter and drives Fermi acceleration process, until it becomes microscopic scales 
enough for particles to escape from trapping.}
\end{figure}

In this paper, we propose acceleration model during the dynamic process of interaction between bi-directionally ejected fractal plasmoids and fast shocks just below and 
above the current sheet. Here we focus on the fractal distribution of plasmoids and their role for the trapping favorable for acceleration.\\

{\it Scenario of First-order Fermi Acceleration} --
We propose a Fermi acceleration process when multiple plasmoids collide with a fast shock above the loop-top. Once reconnection occurs in the corona, reconnection 
outflow generates termination shocks (fast shocks) above the loop-top and below the magnetic flux rope \cite{yok01}. The scenario we propose is as follows: (1) multiple 
plasmoids formed in a fractal current sheet are ejected downward (upward) and collide with a fast shock below (above) the current sheet with the speed less than or comparable 
to the Alfv\'{e}n velocity (300-1000 km/s) with trapped particles inside (Fig. 1). (2) During the collision, magnetic pressure gradient at the shock front, i.e. magnetic mirror force, 
reflects and traps particles upstream the fast shock (Fig. 2). (3) Through the passage of the plasmoid, the distance between two reflection points along the field line, $L$, becomes 
shorter and shorter. At that time, each particle gets momentum from the shock front $\Delta$($mv$)=2$m u$ via one collision, resulting into the first-order Fermi acceleration. 
Here $m$ and $u$ are particle mass and the relative velocity of the shock front to the rest frame of the upstream. (4) Finally when the bouncing timescale 
becomes comparable to the reflection time scale determined by ion cyclotron frequency (2-10)$\Omega_{ci}^{-1}$ \citep{chi88, kra89}, in other words, when the distance $L$ 
becomes comparable to the Larmor radius, particles escape from the trapping upstream the fast shock.

During this process, plasmoids disturb the ambient plasma and scatter particles. It is also known that the parallel motion of a particle tends to change into the perpendicular 
one at the rotational discontinuity. The same process may happen when magnetic field lines are bent at the fast shocks. Since the pitch angle increases during the passage of 
the shock front, it would positively work for the trapping of particles. Similarly, betatron acceleration by compression of a plasmoid downstream and even 
upstream the shock will play the same role in continuing reflection longer. 
The guide field enables particles to move parallel to the shock and further DC-field acceleration occurs. 

\begin{figure}[b]
\includegraphics{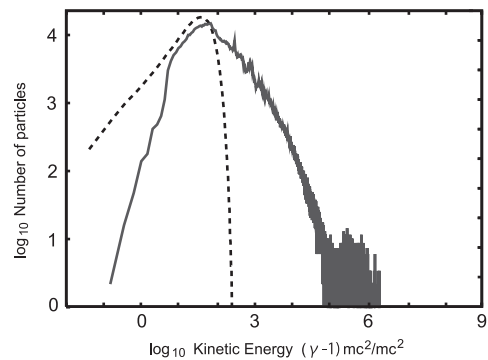}
\caption{\label{fig3} Particle energy spectrum calculated by solving equations of motion of test particles, where dotted and solid lines show energy spectra before and after 
acceleration, respectively. We assumed a cylindrical flux tube and compressed it below the shock front with the compression ratio less than 4 moving upward with 
Alfv\'{e}n velocity in the rest frame of the upstream, in which particles are trapped and accelerated by the electric field inside the shock front of finite width.}
\end{figure}

For the first-order Fermi process, energy gain of particles is described by dE/dt = 4uE/L \cite{fer49}. Particles conserve the action parallel to the field lines, so that the 
time variation of particle energy is E(t) = C/L$^2$ =C/(L$_0$-2ut)$^2$, where C is constant. This leads to the ratio of accelerated and injected particle energies written 
by the ratio of Larmor radius of accelerated particles r$_{L,acc}$ and the size of a plasmoid L$_0$, 
\begin{equation}
\frac{E_{acc}}{E_0} = \left({\frac{L_0}{r_{L,acc}}}\right)^2 
= 10^5 \left({\frac{L_0}{10^4 {\rm m}}}\right)^2 \left({\frac{r_{L,acc}}{30 {\rm m}}}\right)^{-2}.
\end{equation}
Here the adiabatic invariant leads to $E_0L_0^2$= $E_{acc}r_{L,acc}^2$= $E_{acc}(\gamma mc/eB)^2$=$E_{acc}(E_{acc}/eBc)^2$=$E_{acc}^3/(eBc)^2$. Then, 
we finally get the following equation,
\begin{equation}
\frac{E_{acc}}{E_0} = 4640 \left({\frac{L_0}{10^4 {\rm m}}}\right)^{\frac{2}{3}} \left({\frac{B}{10 {\rm G}}}\right)^{\frac{2}{3}} \left({\frac{E_0}{10 {\rm keV}}}\right)^{-\frac{2}{3}}
\end{equation}

Here we note that Larmor radius in the corona is 1-100 m, so that the scale gap between Larmor radius and the size of a plasmoid (10$^4$-10$^{8}$ m) 
quite positively works on this acceleration process. If particles fall into the loss cone before the end of the bounce motion, the energy gain would be determined by the final 
loop length. The time scale of acceleration corresponds to the transit time of a plasmoid through the shock front,
\begin{equation}
\tau_{acc} = \frac{L_0}{v_A} = 10 \left({\frac{L_0}{10^4 {\rm m}}}\right) \left({\frac{v_A}{1000 {\rm km/s}}}\right)^{-1} {\rm [ms]}
\end{equation}
This means effective acceleration in a short time period.

This acceleration process is without the selection of protons and electrons. Electrons gain relatively small energy compared with protons but complement by the number 
of collisions. The injection energy necessary for this acceleration process is determined by the two constraints. One is that initial particle velocity is super-Alfv\'{e}nic, 
when a trapped particle can repeat bouncing upstream the fast shock. The other one is that accelerated particles overcome the energy loss rate by Coulomb collision 
$k_B T/\tau_{ei}$, where $k_B$ is Boltzmann constant and $\tau_{ei}$ is collision time between electrons and protons such that 
$\frac{dE}{dt} = \frac{uE}{L} -4.9\times 10^{-9} \left({ \frac{n}{{\rm cm^{-3}}} }\right) \left({ \frac{E}{{\rm keV}} }\right)^{-\frac{1}{2}} {\rm [keV/s]} \ge 0$
for electrons, where $E$ is electron energy and $n$ is electron density. This constraint leads to the lower cut off energy of electrons, 
$E_c=13 \left({ \frac{n}{10^{10} {\rm cm^{-3}}} }\right)^{\frac{2}{3}} \left({ \frac{L}{u}} \right)^{\frac{2}{3}} {\rm [keV]}$
\cite{tsu98}. For protons, 
$E_c=0.11 \left({ \frac{n}{10^{10} {\rm cm^{-3}}} }\right)^{\frac{2}{3}} \left({ \frac{L}{u} }\right)^{\frac{2}{3}} {\rm [MeV]}$.
These constraints are greater than super-Alfv\'{e}nic and require additional acceleration in the initial phase. Electrons are heated up by slow shocks 
elongating from the reconnection point, from $E_0$ to $E_0$/$\beta$ where $\beta$ is plasma beta ($\sim$0.01 in the corona). On the other hand, protons could be 
accelerated in multiple X-points via merging of plasmoids \cite{tan10, oka10, kar11} i.e. $E_0$= $eEd$=$ev_{in}Bd$= 100 keV ($v_{in}$/1 km s$^{-1}$)($B$/100 G)($d$/10 
km) as well as Fermi process in a shrinking plasmoid after the coalescence \cite{dra06, tan10}, where $d$ is acceleration distance and $v_{in}$ is inflow velocity.

It is interesting to note that multiple internal shocks can be generated in reconnection outflows during the nonlinear evolution with finite fluctuation at the diffusion region 
\cite{tan07}. At that time, multiple small-scale plasmoids can interact with the internal shocks in a fractal manner as well as large plasmoids do with the loop-top shocks.\\
%

{\it Integrated Power-law Spectrum} --
The energy spectrum of particles accelerated in a plasmoid colliding with a fast shock is derived from the equation of mass conservation in energy space,
\begin{equation}
\frac{\partial N}{\partial t}+\frac{\partial }{\partial E} \left({\frac{\partial E}{\partial t} N(E,t)}\right) =-\frac{N}{\tau}
\end{equation}
where $\tau$ is the run-away time scale much larger than the acceleration time $t_{acc}$ and hereafter we neglect the first term $\partial N/\partial t$ (if we consider 
time dependence of $N(E,t)$, a factor $\exp$(-$\alpha$t/$\tau$) is multiplied to the steady solution). For the first-order Fermi process, $dE/dt = 4uE/L = 4uE^{3/2}$
$/C^{1/2}$ because EL$^2$=C. If we assume the relative shock velocity along the field line $u$ is independent of particle energy $E$, we get the solution
\begin{eqnarray}
N(E) = E^{-3/2} \exp \left({\frac{\sqrt{C}}{2u\tau \sqrt{E}}}\right) \propto E^{-3/2}.
\end{eqnarray}

However, the relative shock velocity $u$ actually depends on the distance of bounce motion along the closed field lines between the two reflection points of a plasmoid 
$L(E)$, such that $u(E)$ = $u_0/\sin \frac{\theta}{2}$ = $u_0/\sin (L/2R)$= $u_0/\sin (\sqrt{C}/2R\sqrt{E})$, where $u_0$ is the relative velocity of the shock front normal 
to the rest frame of upstream ($\sim$Alfv\'{e}n velocity) and $\theta$ is the central angle of the curvature $L$ ($\theta$=$L/R$; $R$ is the radius of curvature). 
Then we get the solution,
\begin{eqnarray}
N(E) = E^{-3/2} \exp \left({-\frac{\sqrt{C}}{2u\tau E^{\frac{3}{2}}} -\int \frac{du/dE}{u} dE }\right). 
\end{eqnarray}
\begin{figure}[htb]
\includegraphics{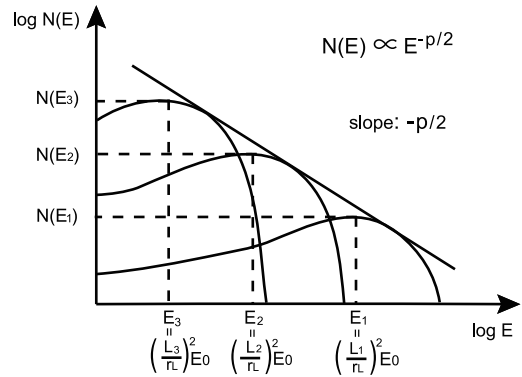}
\caption{\label{fig4} Superposed power-law distribution of multi energy spectra of plasmoids colliding with the fast shock.}
\end{figure}
In an extreme case ($\theta \ll$1), the relative shock velocity is u(E) = 2$u_0/\theta$ =2$u_0 R/L$ =2$\frac{Ru_0}{\sqrt{C}} \sqrt{E}$, therefore, $du/dE = u/2E$. 
This simplifies the equation (6) to 
\begin{eqnarray}
N(E) = E^{-2} \exp \left({\frac{C}{8u_0\tau RE}}\right) \propto E^{-2}.
\end{eqnarray}
which asymptotically approaches to $E^{-2}$ for $E \gg$1.

For comparison, we performed test particle simulation. We assumed 2D magnetic field configuration of a closed magnetic island colliding with a fast shock with analytical 
model whose downstream side is constantly compressed in the rest frame of a plasmoid, in which trajectories and energies of protons were calculated by solving gyromotion. 
The initial energy spectrum is soft power-law distribution with power-law index 7 (Fig. 3). This numerical simulation shows that particles are effectively accelerated at the 
shock front and the power-law spectrum grows harder to the power-law index -2. This is consistent with the previous estimation (detailed discussion in Nishizuka et al. 
in prep).

The observed HXR spectrum would be the superposition of several HXR spectra emitted from the numbers of plasmoids colliding with a fast shock. Both accelerated 
particle energy and acceleration time are proportional to the size of a plasmoid L$_0$, as shown in equations (1)-(3). Therefore, if the size of a plasmoid $L_0$ depends 
on the power-law distribution via fractal reconnection process, the power-law distribution of the observed HXR spectrum can be naturally explained, even if some energy 
spectra from plasmoids does not develop to the power-law distribution enough. Each spectrum tends to have the peak at the temperature determined by the adiabatic 
heating, i.e. $E=(L_0/r_L)^2E_0$. Then, here we assume the power-law distribution of plasmoids with power-law index $\alpha$, i.e. $N(L_0)\propto L_0^{-\alpha}$ (fractal 
current sheet), and the power-law energy spectrum with power-law index $p$, $N(E)$ $\propto$ E$^{-p}$. At that time, these two power-law indices $\alpha$ and $p$ are 
related by the equation $N(L_0)dL_0=N(E)dE$, simplified to $p=\frac{1}{2}(\alpha+1)$. With numerical simulation, the power-law index $\alpha$ is shown to be $\alpha$=2 
\cite{lou11, nis09}.  If we adopt $\alpha=2$ based on the simulation result, this equation gives us $p=\frac{3}{2}$, which is harder than the observation, though it can 
become softer if $\alpha$ varies. This may also indicate the possibility that we can expect microscopic parameter $\alpha$ from the observed HXR spectrum.

Our model is a unified model of magnetic reconnection and particle acceleration in a solar flare, in which particles are effectively accelerated coupled with the dynamics 
of plasmoid ejections colliding with fast shocks. Impulsive ejections of fractal plasmoids explain observed short time variations of HXR and $\gamma$-ray bursts ($<$1 s), 
HXR power-law spectrum and maximum energy of acceleration (MeV electrons and GeV protons). Multiple ejections of plasmoids also generate turbulent outflows and 
internal multiple shocks \cite{tan07}. These would further enable stochastic acceleration. The trapping between fast shocks and plasmoids increases local coronal 
density and may explain observed HXR flux via bremsstrahlung. Our model can be  applied to fast shocks both at the loop-top and below the flux rope, which is consistent 
with double coronal HXR sources, and then also applied to astrophysical jets with high energy particles.\\

We acknowledge the supports in part by a JSPS Research Grant, in part by the JSPS Core-to-Core Program 22001 and by the Grant-in-Aid for the 
Global COE Program ``The Next Generation of Physics, Spun from Universality and Emergence''. The numerical simulation was carried out on Altix2700 at YITP in 
Kyoto University.

\bibliography{apssamp110703f_}

\providecommand{\noopsort}[1]{}\providecommand{\singleletter}[1]{#1}%
\begin{thebibliography}{10}%
\makeatletter
\providecommand \@ifxundefined [1]{%
 \ifx #1\undefined \expandafter \@firstoftwo
 \else \expandafter \@secondoftwo
\fi
}%
\providecommand \@ifnum [1]{%
 \ifnum #1\expandafter \@firstoftwo
 \else \expandafter \@secondoftwo
\fi
}%
\providecommand \enquote [1]{``#1''}%
\providecommand \bibnamefont  [1]{#1}%
\providecommand \bibfnamefont [1]{#1}%
\providecommand \citenamefont [1]{#1}%
\providecommand\href[0]{\@sanitize\@href}%
\providecommand\@href[1]{\endgroup\@@startlink{#1}\endgroup\@@href}%
\providecommand\@@href[1]{#1\@@endlink}%
\providecommand \@sanitize [0]{\begingroup\catcode`\&12\catcode`\#12\relax}%
\@ifxundefined \pdfoutput {\@firstoftwo}{%
 \@ifnum{\z@=\pdfoutput}{\@firstoftwo}{\@secondoftwo}%
}{%
 \providecommand\@@startlink[1]{\leavevmode\special{html:<a href="#1">}}%
 \providecommand\@@endlink[0]{\special{html:</a>}}%
}{%
 \providecommand\@@startlink[1]{%
  \leavevmode
  \pdfstartlink
   attr{/Border[0 0 1 ]/H/I/C[0 1 1]}%
   user{/Subtype/Link/A<</Type/Action/S/URI/URI(#1)>>}%
  \relax
 }%
 \providecommand\@@endlink[0]{\pdfendlink}%
}%
\providecommand \url  [0]{\begingroup\@sanitize \@url }%
\providecommand \@url [1]{\endgroup\@href {#1}{\urlprefix}}%
\providecommand \urlprefix [0]{URL }%
\providecommand \Eprint[0]{\href }%
\@ifxundefined \urlstyle {%
  \providecommand \doi [1]{doi:\discretionary{}{}{}#1}%
}{%
  \providecommand \doi [0]{doi:\discretionary{}{}{}\begingroup
  \urlstyle{rm}\Url }%
}%
\providecommand \doibase [0]{http://dx.doi.org/}%
\providecommand \Doi[1]{\href{\doibase#1}}%
\providecommand \bibAnnote [3]{%
  \BibitemShut{#1}%
  \begin{quotation}\noindent
    \textsc{Key:}\ #2\\\textsc{Annotation:}\ #3%
  \end{quotation}%
}%
\providecommand \bibAnnoteFile [2]{%
  \IfFileExists{#2}{\bibAnnote {#1} {#2} {\input{#2}}}{}%
}%
\providecommand \typeout [0]{\immediate \write \m@ne }%
\providecommand \selectlanguage [0]{\@gobble}%
\providecommand \bibinfo [0]{\@secondoftwo}%
\providecommand \bibfield [0]{\@secondoftwo}%
\providecommand \translation [1]{[#1]}%
\providecommand \BibitemOpen[0]{}%
\providecommand \bibitemStop [0]{}%
\providecommand \bibitemNoStop [0]{.\EOS\space}%
\providecommand \EOS [0]{\spacefactor3000\relax}%
\providecommand \BibitemShut [1]{\csname bibitem#1\endcsname}%
\bibitem{asc01}%
  \BibitemOpen
  \bibfield{author}{%
  \bibinfo {author} {\bibfnamefont{M.~J.}\ \bibnamefont{Aschwanden}},\ }%
  \emph{\bibinfo {title} {Particle acceleration and kinematics in solar
  flares}}\ (\bibinfo {publisher} {Kluwer Academic Publishers},\ \bibinfo
  {year} {2001})%
  \bibAnnoteFile{NoStop}{asc01}%
\bibitem{mas94}%
  \BibitemOpen
  \bibfield{author}{%
  \bibinfo {author} {\bibfnamefont{S.}~\bibnamefont{Masuda}} \emph{et~al.},\ }%
  \bibfield{journal}{%
  \bibinfo {journal} {Nature}\ }%
  \textbf{\bibinfo {volume} {371}},\ \bibinfo {pages} {495} (\bibinfo {year}
  {1994})%
  \bibAnnoteFile{NoStop}{mas94}%
\bibitem{asc96}%
  \BibitemOpen
  \bibfield{author}{%
  \bibinfo {author} {\bibfnamefont{M.~J.}\ \bibnamefont{Aschwanden}}, \bibinfo
  {author} {\bibfnamefont{H.}~\bibnamefont{Hudson}}, \bibinfo {author}
  {\bibfnamefont{T.}~\bibnamefont{Kosugi}},\ and\ \bibinfo {author}
  {\bibfnamefont{R.~A.}\ \bibnamefont{Schwartz}},\ }%
  \bibfield{journal}{%
  \bibinfo {journal} {Astrophys. J.}\ }%
  \textbf{\bibinfo {volume} {464}},\ \bibinfo {pages} {985} (\bibinfo {year}
  {1996})%
  \bibAnnoteFile{NoStop}{asc96}%
\bibitem{lin71}%
  \BibitemOpen
  \bibfield{author}{%
  \bibinfo {author} {\bibfnamefont{R.~P.}\ \bibnamefont{Lin}}\ and\ \bibinfo
  {author} {\bibfnamefont{H.~S.}\ \bibnamefont{Hudson}},\ }%
  \bibfield{journal}{%
  \bibinfo {journal} {Sol. Phys.}\ }%
  \textbf{\bibinfo {volume} {17}},\ \bibinfo {pages} {412} (\bibinfo {year}
  {1971})%
  \bibAnnoteFile{NoStop}{lin71}%
\bibitem{hol85}%
  \BibitemOpen
  \bibfield{author}{%
  \bibinfo {author} {\bibfnamefont{G.~D.}\ \bibnamefont{Holman}},\ }%
  \bibfield{journal}{%
  \bibinfo {journal} {Astrophys. J.}\ }%
  \textbf{\bibinfo {volume} {293}},\ \bibinfo {pages} {584} (\bibinfo {year}
  {1985})%
  \bibAnnoteFile{NoStop}{hol85}%
\bibitem{ben94}%
  \BibitemOpen
  \bibfield{author}{%
  \bibinfo {author} {\bibfnamefont{S.~G.}\ \bibnamefont{Benka}}\ and\ \bibinfo
  {author} {\bibfnamefont{G.~D.}\ \bibnamefont{Holman}},\ }%
  \bibfield{journal}{%
  \bibinfo {journal} {Astrophys. J.}\ }%
  \textbf{\bibinfo {volume} {435}},\ \bibinfo {pages} {469} (\bibinfo {year}
  {1994})%
  \bibAnnoteFile{NoStop}{ben94}%
\bibitem{lit96}%
  \BibitemOpen
  \bibfield{author}{%
  \bibinfo {author} {\bibfnamefont{Y.~E.}\ \bibnamefont{Litvinenko}},\ }%
  \bibfield{journal}{%
  \bibinfo {journal} {Astrophys. J.}\ }%
  \textbf{\bibinfo {volume} {997}},\ \bibinfo {pages} {462} (\bibinfo {year}
  {1996})%
  \bibAnnoteFile{NoStop}{lit96}%
\bibitem{bro85}%
  \BibitemOpen
  \bibfield{author}{%
  \bibinfo {author} {\bibfnamefont{J.~C.}\ \bibnamefont{Brown}}\ and\ \bibinfo
  {author} {\bibfnamefont{J.~M.}\ \bibnamefont{Loran}},\ }%
  \bibfield{journal}{%
  \bibinfo {journal} {Mon. Not. R. Astron. Soc.}\ }%
  \textbf{\bibinfo {volume} {212}},\ \bibinfo {pages} {245} (\bibinfo {year}
  {1985})%
  \bibAnnoteFile{NoStop}{bro85}%
\bibitem{ben87}%
  \BibitemOpen
  \bibfield{author}{%
  \bibinfo {author} {\bibfnamefont{A.~O.}\ \bibnamefont{Benz}}\ and\ \bibinfo
  {author} {\bibfnamefont{D.~F.}\ \bibnamefont{Smith}},\ }%
  \bibfield{journal}{%
  \bibinfo {journal} {Sol. Phys.}\ }%
  \textbf{\bibinfo {volume} {107}},\ \bibinfo {pages} {299} (\bibinfo {year}
  {1987})%
  \bibAnnoteFile{NoStop}{ben87}%
\bibitem{ter89}%
  \BibitemOpen
  \bibfield{author}{%
  \bibinfo {author} {\bibfnamefont{T.}~\bibnamefont{Terasawa}}\ and\ \bibinfo
  {author} {\bibfnamefont{M.}~\bibnamefont{Scholer}},\ }%
  \bibfield{journal}{%
  \bibinfo {journal} {Science}\ }%
  \textbf{\bibinfo {volume} {244}},\ \bibinfo {pages} {1050} (\bibinfo {year}
  {1989})%
  \bibAnnoteFile{NoStop}{ter89}%
\bibitem{mil96}%
  \BibitemOpen
  \bibfield{author}{%
  \bibinfo {author} {\bibfnamefont{J.~A.}\ \bibnamefont{Miller}}, \bibinfo
  {author} {\bibfnamefont{T.~N.}\ \bibnamefont{LaRosa}},\ and\ \bibinfo
  {author} {\bibfnamefont{R.~L.}\ \bibnamefont{Moore}},\ }%
  \bibfield{journal}{%
  \bibinfo {journal} {Astrophys. J.}\ }%
  \textbf{\bibinfo {volume} {461}},\ \bibinfo {pages} {445} (\bibinfo {year}
  {1996})%
  \bibAnnoteFile{NoStop}{mil96}%
\bibitem{som97}%
  \BibitemOpen
  \bibfield{author}{%
  \bibinfo {author} {\bibfnamefont{B.~V.}\ \bibnamefont{Somov}}\ and\ \bibinfo
  {author} {\bibfnamefont{T.}~\bibnamefont{Kosugi}},\ }%
  \bibfield{journal}{%
  \bibinfo {journal} {Astrophys. J.}\ }%
  \textbf{\bibinfo {volume} {485}},\ \bibinfo {pages} {859} (\bibinfo {year}
  {1997})%
  \bibAnnoteFile{NoStop}{som97}%
\bibitem{tsu98}%
  \BibitemOpen
  \bibfield{author}{%
  \bibinfo {author} {\bibfnamefont{S.}~\bibnamefont{Tsuneta}}\ and\ \bibinfo
  {author} {\bibfnamefont{T.}~\bibnamefont{Naito}},\ }%
  \bibfield{journal}{%
  \bibinfo {journal} {Astrophys. J. Lett.}\ }%
  \textbf{\bibinfo {volume} {495}},\ \bibinfo {pages} {67} (\bibinfo {year}
  {1998})%
  \bibAnnoteFile{NoStop}{tsu98}%
\bibitem{dra06}%
  \BibitemOpen
  \bibfield{author}{%
  \bibinfo {author} {\bibfnamefont{J.~F.}\ \bibnamefont{Drake}}, \bibinfo
  {author} {\bibfnamefont{M.}~\bibnamefont{Swisdak}}, \bibinfo {author}
  {\bibfnamefont{H.}~\bibnamefont{Che}},\ and\ \bibinfo {author}
  {\bibfnamefont{M.~A.}\ \bibnamefont{Shay}},\ }%
  \bibfield{journal}{%
  \bibinfo {journal} {Nature}\ }%
  \textbf{\bibinfo {volume} {433}},\ \bibinfo {pages} {553} (\bibinfo {year}
  {2006})%
  \bibAnnoteFile{NoStop}{dra06}%
\bibitem{dau11}%
  \BibitemOpen
  \bibfield{author}{%
  \bibinfo {author} {\bibfnamefont{W.}~\bibnamefont{Daughton}} \emph{et~al.},\
  }%
  \bibfield{journal}{%
  \bibinfo {journal} {Nature Phys.}\ }%
  \textbf{\bibinfo {volume} {7}},\ \bibinfo {pages} {539} (\bibinfo {year}
  {2011})%
  \bibAnnoteFile{NoStop}{dau11}%
\bibitem{hos12}%
  \BibitemOpen
  \bibfield{author}{%
  \bibinfo {author} {\bibfnamefont{M.}~\bibnamefont{Hoshino}},\ }%
  \bibfield{journal}{%
  \bibinfo {journal} {Phys. Rev. Lett.}\ }%
  \textbf{\bibinfo {volume} {108}},\ \bibinfo {pages} {135003} (\bibinfo {year}
  {2012})%
  \bibAnnoteFile{NoStop}{hos12}%
\bibitem{shi11}%
  \BibitemOpen
  \bibfield{author}{%
  \bibinfo {author} {\bibfnamefont{K.}~\bibnamefont{Shibata}}\ and\ \bibinfo
  {author} {\bibfnamefont{T.}~\bibnamefont{Magara}},\ }%
  \bibfield{journal}{%
  \bibinfo {journal} {Living Review in Solar Phys.}\ }%
  \textbf{\bibinfo {volume} {8}},\ \bibinfo {pages} {6} (\bibinfo {year}
  {2011})%
  \bibAnnoteFile{NoStop}{shi11}%
\bibitem{ohy97}%
  \BibitemOpen
  \bibfield{author}{%
  \bibinfo {author} {\bibfnamefont{M.}~\bibnamefont{Ohyama}}\ and\ \bibinfo
  {author} {\bibfnamefont{K.}~\bibnamefont{Shibata}},\ }%
  \bibfield{journal}{%
  \bibinfo {journal} {Publ. Astron. Soc. Jap.}\ }%
  \textbf{\bibinfo {volume} {49}},\ \bibinfo {pages} {249} (\bibinfo {year}
  {1997})%
  \bibAnnoteFile{NoStop}{ohy97}%
\bibitem{kar04}%
  \BibitemOpen
  \bibfield{author}{%
  \bibinfo {author} {\bibfnamefont{M.}~\bibnamefont{Karlick{\'{y}}}},\ }%
  \bibfield{journal}{%
  \bibinfo {journal} {Astron. Astrophys.}\ }%
  \textbf{\bibinfo {volume} {417}},\ \bibinfo {pages} {325} (\bibinfo {year}
  {2004})%
  \bibAnnoteFile{NoStop}{kar04}%
\bibitem{nis10}%
  \BibitemOpen
  \bibfield{author}{%
  \bibinfo {author} {\bibfnamefont{N.}~\bibnamefont{Nishizuka}}, \bibinfo
  {author} {\bibfnamefont{H.}~\bibnamefont{Takasaki}}, \bibinfo {author}
  {\bibfnamefont{A.}~\bibnamefont{Asai}},\ and\ \bibinfo {author}
  {\bibfnamefont{K.}~\bibnamefont{Shibata}},\ }%
  \bibfield{journal}{%
  \bibinfo {journal} {Astrophys. J.}\ }%
  \textbf{\bibinfo {volume} {711}},\ \bibinfo {pages} {1062} (\bibinfo {year}
  {2010})%
  \bibAnnoteFile{NoStop}{nis10}%
\bibitem{asa04}%
  \BibitemOpen
  \bibfield{author}{%
  \bibinfo {author} {\bibfnamefont{A.}~\bibnamefont{Asai}}, \bibinfo {author}
  {\bibfnamefont{T.}~\bibnamefont{Yokoyama}}, \bibinfo {author}
  {\bibfnamefont{M.}~\bibnamefont{Shimojo}},\ and\ \bibinfo {author}
  {\bibfnamefont{K.}~\bibnamefont{Shibata}},\ }%
  \bibfield{journal}{%
  \bibinfo {journal} {Astrophys. J.}\ }%
  \textbf{\bibinfo {volume} {605}},\ \bibinfo {pages} {77} (\bibinfo {year}
  {2004})%
  \bibAnnoteFile{NoStop}{asa04}%
\bibitem{kol07}%
  \BibitemOpen
  \bibfield{author}{%
  \bibinfo {author} {\bibfnamefont{S.}~\bibnamefont{Koloma{\'{n}}ski}}\ and\
  \bibinfo {author} {\bibfnamefont{M.}~\bibnamefont{Karlick{\'{y}}}},\ }%
  \bibfield{journal}{%
  \bibinfo {journal} {Astron. Astrophys.}\ }%
  \textbf{\bibinfo {volume} {475}},\ \bibinfo {pages} {685} (\bibinfo {year}
  {2007})%
  \bibAnnoteFile{NoStop}{kol07}%
\bibitem{sui03}%
  \BibitemOpen
  \bibfield{author}{%
  \bibinfo {author} {\bibfnamefont{L.}~\bibnamefont{Sui}}\ and\ \bibinfo
  {author} {\bibfnamefont{G.~D.}\ \bibnamefont{Holman}},\ }%
  \bibfield{journal}{%
  \bibinfo {journal} {Astrophys. J. Lett.}\ }%
  \textbf{\bibinfo {volume} {251}},\ \bibinfo {pages} {596} (\bibinfo {year}
  {2003})%
  \bibAnnoteFile{NoStop}{sui03}%
\bibitem{shi01}%
  \BibitemOpen
  \bibfield{author}{%
  \bibinfo {author} {\bibfnamefont{K.}~\bibnamefont{Shibata}}\ and\ \bibinfo
  {author} {\bibfnamefont{S.}~\bibnamefont{Tanuma}},\ }%
  \bibfield{journal}{%
  \bibinfo {journal} {Earth, Planets and Space}\ }%
  \textbf{\bibinfo {volume} {53}},\ \bibinfo {pages} {473} (\bibinfo {year}
  {2001})%
  \bibAnnoteFile{NoStop}{shi01}%
\bibitem{nisd09}%
  \BibitemOpen
  \bibfield{author}{%
  \bibinfo {author} {\bibfnamefont{K.}~\bibnamefont{Nishida}} \emph{et~al.},\
  }%
  \bibfield{journal}{%
  \bibinfo {journal} {Astrophys. J.}\ }%
  \textbf{\bibinfo {volume} {690}},\ \bibinfo {pages} {748} (\bibinfo {year}
  {2009})%
  \bibAnnoteFile{NoStop}{nisd09}%
\bibitem{hua10}%
  \BibitemOpen
  \bibfield{author}{%
  \bibinfo {author} {\bibfnamefont{Y.~M.}\ \bibnamefont{Huang}}\ and\ \bibinfo
  {author} {\bibfnamefont{A.}~\bibnamefont{Bhattacharjee}},\ }%
  \bibfield{journal}{%
  \bibinfo {journal} {Phys. Plasmas}\ }%
  \textbf{\bibinfo {volume} {17}},\ \bibinfo {pages} {062104} (\bibinfo {year}
  {2010})%
  \bibAnnoteFile{NoStop}{hua10}%
\bibitem{lou09}%
  \BibitemOpen
  \bibfield{author}{%
  \bibinfo {author} {\bibfnamefont{N.~F.}\ \bibnamefont{Loureiro}}
  \emph{et~al.},\ }%
  \bibfield{journal}{%
  \bibinfo {journal} {Mon. Not. R. Astron. Soc.}\ }%
  \textbf{\bibinfo {volume} {399}},\ \bibinfo {pages} {146} (\bibinfo {year}
  {2009})%
  \bibAnnoteFile{NoStop}{lou09}%
\bibitem{sam09}%
  \BibitemOpen
  \bibfield{author}{%
  \bibinfo {author} {\bibfnamefont{R.}~\bibnamefont{Samtaney}} \emph{et~al.},\
  }%
  \bibfield{journal}{%
  \bibinfo {journal} {Phys. Rev. Lett.}\ }%
  \textbf{\bibinfo {volume} {103}},\ \bibinfo {pages} {105004} (\bibinfo {year}
  {2009})%
  \bibAnnoteFile{NoStop}{sam09}%
\bibitem{bar11}%
  \BibitemOpen
  \bibfield{author}{%
  \bibinfo {author} {\bibfnamefont{M.}~\bibnamefont{B{\'{a}}rta}}, \bibinfo
  {author} {\bibfnamefont{J.}~\bibnamefont{B{\"{u}}chner}}, \bibinfo {author}
  {\bibfnamefont{M.}~\bibnamefont{Karlick{\'{y}}}},\ and\ \bibinfo {author}
  {\bibfnamefont{P.}~\bibnamefont{Kotr{\v{c}}}},\ }%
  \bibfield{journal}{%
  \bibinfo {journal} {Astrophys. J.}\ }%
  \textbf{\bibinfo {volume} {730}},\ \bibinfo {pages} {47} (\bibinfo {year}
  {2011})%
  \bibAnnoteFile{NoStop}{bar11}%
\bibitem{kow11}%
  \BibitemOpen
  \bibfield{author}{%
  \bibinfo {author} {\bibfnamefont{G.}~\bibnamefont{Kowal}}, \bibinfo {author}
  {\bibfnamefont{E.~M.}\ \bibnamefont{de~Gouveia Dal~Pino}},\ and\ \bibinfo
  {author} {\bibfnamefont{A.}~\bibnamefont{Lazarian}},\ }%
  \bibfield{journal}{%
  \bibinfo {journal} {Astrophys. J.}\ }%
  \textbf{\bibinfo {volume} {735}},\ \bibinfo {pages} {102} (\bibinfo {year}
  {2011})%
  \bibAnnoteFile{NoStop}{kow11}%
\bibitem{lou11}%
  \BibitemOpen
  \bibfield{author}{%
  \bibinfo {author} {\bibfnamefont{N.~F.}\ \bibnamefont{Loureiro}}, \bibinfo
  {author} {\bibfnamefont{R.}~\bibnamefont{Samtaney}}, \bibinfo {author}
  {\bibfnamefont{A.~A.}\ \bibnamefont{Schekochihin}},\ and\ \bibinfo {author}
  {\bibfnamefont{D.~A.}\ \bibnamefont{Uzdensky}},\ }%
  \bibfield{journal}{%
  \bibinfo {journal} {Phys. Rev. Let.}\ }%
  \textbf{\bibinfo {volume} {11}},\ \bibinfo {pages} {111} (\bibinfo {year}
  {2011})%
  \bibAnnoteFile{NoStop}{lou11}%
\bibitem{she11}%
  \BibitemOpen
  \bibfield{author}{%
  \bibinfo {author} {\bibfnamefont{C.}~\bibnamefont{Shen}}, \bibinfo {author}
  {\bibfnamefont{J.}~\bibnamefont{Lin}},\ and\ \bibinfo {author}
  {\bibfnamefont{N.~A.}\ \bibnamefont{Murphy}},\ }%
  \bibfield{journal}{%
  \bibinfo {journal} {Astrophys. J.}\ }%
  \textbf{\bibinfo {volume} {737}},\ \bibinfo {pages} {14} (\bibinfo {year}
  {2011})%
  \bibAnnoteFile{NoStop}{she11}%
\bibitem{taj87}%
  \BibitemOpen
  \bibfield{author}{%
  \bibinfo {author} {\bibfnamefont{T.}~\bibnamefont{Tajima}} \emph{et~al.},\ }%
  \bibfield{journal}{%
  \bibinfo {journal} {Astrophys. J.}\ }%
  \textbf{\bibinfo {volume} {321}},\ \bibinfo {pages} {1031} (\bibinfo {year}
  {1987})%
  \bibAnnoteFile{NoStop}{taj87}%
\bibitem{oka10}%
  \BibitemOpen
  \bibfield{author}{%
  \bibinfo {author} {\bibfnamefont{M.}~\bibnamefont{Oka}} \emph{et~al.},\ }%
  \bibfield{journal}{%
  \bibinfo {journal} {Astrophys. J.}\ }%
  \textbf{\bibinfo {volume} {714}},\ \bibinfo {pages} {915} (\bibinfo {year}
  {2010})%
  \bibAnnoteFile{NoStop}{oka10}%
\bibitem{tan10}%
  \BibitemOpen
  \bibfield{author}{%
  \bibinfo {author} {\bibfnamefont{K.~G.}\ \bibnamefont{Tanaka}}
  \emph{et~al.},\ }%
  \bibfield{journal}{%
  \bibinfo {journal} {Phys. Plasmas}\ }%
  \textbf{\bibinfo {volume} {17}},\ \bibinfo {pages} {102902} (\bibinfo {year}
  {2010})%
  \bibAnnoteFile{NoStop}{tan10}%
\bibitem{kar11}%
  \BibitemOpen
  \bibfield{author}{%
  \bibinfo {author} {\bibfnamefont{M.}~\bibnamefont{Karlick{\'{y}}}}\ and\
  \bibinfo {author} {\bibfnamefont{M.}~\bibnamefont{Barta}},\ }%
  \bibfield{journal}{%
  \bibinfo {journal} {Astrophys. J.}\ }%
  \textbf{\bibinfo {volume} {733}},\ \bibinfo {pages} {107} (\bibinfo {year}
  {2011})%
  \bibAnnoteFile{NoStop}{kar11}%
\bibitem{nis09}%
  \BibitemOpen
  \bibfield{author}{%
  \bibinfo {author} {\bibfnamefont{N.}~\bibnamefont{Nishizuka}} \emph{et~al.},\
  }%
  \bibfield{journal}{%
  \bibinfo {journal} {Astrophys. J. Lett.}\ }%
  \textbf{\bibinfo {volume} {694}},\ \bibinfo {pages} {74} (\bibinfo {year}
  {2009})%
  \bibAnnoteFile{NoStop}{nis09}%
\bibitem{yok01}%
  \BibitemOpen
  \bibfield{author}{%
  \bibinfo {author} {\bibfnamefont{T.}~\bibnamefont{Yokoyama}}\ and\ \bibinfo
  {author} {\bibfnamefont{K.}~\bibnamefont{Shibata}},\ }%
  \bibfield{journal}{%
  \bibinfo {journal} {Astrophys. J.}\ }%
  \textbf{\bibinfo {volume} {549}},\ \bibinfo {pages} {1160} (\bibinfo {year}
  {2001})%
  \bibAnnoteFile{NoStop}{yok01}%
\bibitem{chi88}%
  \BibitemOpen
  \bibfield{author}{%
  \bibinfo {author} {\bibfnamefont{T.}~\bibnamefont{Chiueh}},\ }%
  \bibfield{journal}{%
  \bibinfo {journal} {Astrophys. J.}\ }%
  \textbf{\bibinfo {volume} {333}},\ \bibinfo {pages} {366} (\bibinfo {year}
  {1988})%
  \bibAnnoteFile{NoStop}{chi88}%
\bibitem{kra89}%
  \BibitemOpen
  \bibfield{author}{%
  \bibinfo {author} {\bibfnamefont{D.}~\bibnamefont{Krauss-Varban}}\ and\
  \bibinfo {author} {\bibfnamefont{C.~S.}\ \bibnamefont{Wu}},\ }%
  \bibfield{journal}{%
  \bibinfo {journal} {J. Geophys. Res.}\ }%
  \textbf{\bibinfo {volume} {94}},\ \bibinfo {pages} {15367} (\bibinfo {year}
  {1989})%
  \bibAnnoteFile{NoStop}{kra89}%
\bibitem{fer49}%
  \BibitemOpen
  \bibfield{author}{%
  \bibinfo {author} {\bibfnamefont{E.}~\bibnamefont{Fermi}},\ }%
  \bibfield{journal}{%
  \bibinfo {journal} {Phys. Rev.}\ }%
  \textbf{\bibinfo {volume} {75}},\ \bibinfo {pages} {1169} (\bibinfo {year}
  {1949})%
  \bibAnnoteFile{NoStop}{fer49}%
\bibitem{tan07}%
  \BibitemOpen
  \bibfield{author}{%
  \bibinfo {author} {\bibfnamefont{S.}~\bibnamefont{Tanuma}}\ and\ \bibinfo
  {author} {\bibfnamefont{K.}~\bibnamefont{Shibata}},\ }%
  \bibfield{journal}{%
  \bibinfo {journal} {Publ. Astron. Soc. Jap. Lett.}\ }%
  \textbf{\bibinfo {volume} {59}},\ \bibinfo {pages} {1} (\bibinfo {year}
  {2007})%
  \bibAnnoteFile{NoStop}{tan07}%
\bibitem{REVTEX41Control}%
  \BibitemOpen
  %
  \bibAnnoteFile{NoStop}{REVTEX41Control}%
\end{thebibliography}%







\end{document}